\documentclass[preprint,showpacs,nofootinbib,prd,aps]{revtex4-1}
\usepackage{graphicx,amstext,amssymb}

\begin{document}
\title{$K^*(892)$ and $\phi(1020)$ production and their decay into the hadronic medium at
  the Large Hadron Collider} 

\author{V.~M.~Shapoval$^1$}
\author{P.~Braun-Munzinger$^{2}$}
\author{Yu.~M.~Sinyukov$^{1,2}$}
\affiliation{$^1$Bogolyubov Institute for Theoretical Physics,
Metrolohichna  14b, 03680 Kiev,  Ukraine\\
 $^2$ExtreMe Matter Institute EMMI, GSI~Helmholtzzentrum f\"ur~Schwerionenforschung,
D-64291 Darmstadt, Germany}

\begin{abstract}
The production of the $K^*(892)$ strange resonance in Pb+Pb collisions
at $\sqrt{s_{NN}}=2.76$~TeV LHC energy is analyzed within the
integrated hydrokinetic model (iHKM) at different equations of state of superdense matter. The similar analysis is done also for the RHIC top energy $\sqrt{s_{NN}}=200$~GeV for comparison purposes.  A modification of experimental
$K^*(892)$-identification is studied for different centralities in
view of possible re-scattering of the decay products at the
afterburner stage of the fireball evolution. We see quite intensive rescattering of the decay products as well as recombination processes for $K^*(892)$.   In addition, the
production of the much longer-long-lived $\phi(1020)$ resonance with
hidden strange quark content is investigated.
\end{abstract}

\maketitle



\section{Introduction}

The analysis of strange resonance yields in relativistic heavy ion
collisions can provide valuable information about the properties of
hot and dense matter, formed in such processes. The results of
theoretical lattice QCD calculations as well as a multitude of
experimental observations on hadron production, hydrodynamic flow and
jet quenching in ultra-relativistic nuclear collisions demonstrate the
presence of quark-gluon matter in the hot fireball as well as the
existence of a cross-over phase transition from the strongly coupled quark-gluon
plasma (QGP) to hadronic matter. For recent reviews see
~\cite{pbm,schukraft} and references there.

Here we don't want to deal with a description of the QGP phase but
rather concentrate on getting a better handle on the hadronic phase
after chemical freeze-out. 
The hadrons containing strange quarks
can play an important role in understanding inter alia the 
final hadronic stage of the matter evolution in $A+A$ collisions. 
At LHC energy, the chemical freeze-out temperature $T_{chem}$ is very well
determined to be $156.5 \pm 1.5$ MeV
\cite{stachel-sqm2013}.

This temperature is very close to the best value for the
pseudo-critical temperature $T_{ps}$ of $154 \pm
9$~MeV~\cite{lattice-hotqcd,lattice-wuppertal}.
Consequently, we conclude that the non-equilibrium, dilute hadronic
stage commences for temperatures below $T_{ps}$, i.e. the system reaches there
the final
afterburner stage.  At this non-equilibrium stage particles still
collide, at least, elastically, the resonances decay, and hadrons {\it
  gradually} escape from the system and travel freely to the
detectors. Strange hadron resonance states, such as the  $K(892)^*$, that
lives about 4~fm/$c$, can be used as a probe of the intensity of
collision processes at this afterburner stage if one studies their
yield by registering the hadronic decay channel $K(892)^* \rightarrow
K\pi$. Taking into account the $K^*$ lifetime, such information about the medium 
will deal mostly with the first $4-5$~fm/$c$ of the hadronic phase.
The interactions of daughter hadrons may prevent
identification of all produced $K^*$'s in the experiment. Consequently
one expects for such resonances deviations from the overall chemical
freeze-out picture determined using hadrons which are stable against
strong decays. Such deviations are indeed observed~\cite{stachel-sqm2013}.

Dealing with the modification of resonance yields due to modifications in the non-equilibrium hadronic phase after 
chemical freeze-out is a delicate matter. Processes which can lead to $K^*$ losses involve:

1) elastic scattering of the decay pions and kaons with the constituents of the medium, in the present case mostly 
pions and kaons as well as some nucleons. A prominent case is re-formation of the $K^*$. Since the $K^*$ is a 
pseudo-vector, the decay angular distribution in the process should be taken into account. Furthermore, forward 
angle Coulomb scattering should not be neglected, both from other charged mesons and baryons in the fireball. At 
LHC energies, the net charge of the fireball vanishes, contrary to the situation at lower beam energies. However, 
there may be large charges in the causally connected parts of the fireball due to fluctuations and their possible  
influence should be studied. The re-formed $K^*$ may also survive in the medium and decay outside.

2) inelastic processes such as $K^* + \pi \rightarrow K + \pi + \pi$, i.e. processes involving higher $K^*$ 
resonances can also lead to $K^*$ losses.

3) the losses or gains should be confined to the life time of the hadronic phase. At LHC energy the ALICE 
collaboration has investigated this lifetime through measurements of Hanbury-Brown/Twiss interferometry (see~\cite{Aamodt}).
From these measurements one concludes that the life time of the fireball  is rather short, about 10 fm/$c$ \footnote{According to the results of \cite{continuous, mt_scales, Malinina} such a time corresponds to the temporal location of  the maximal emission hypersurface for relatively soft pions. In evolutionary models hadron emission is continuous and typically  has fairly long tail outside the maximum.}.

The studies below are still rather schematic in that the UrQMD model is used to describe the final hadronic 
cascade where all Coulomb effects are neglected and the main collision process which is taken into account is 
$\pi+K \leftrightarrow K^*$. Also the life time of the hadronic phase in UrQMD lasts longer than the 10 fm/$c$    
discussed above. Nevertheless, we believe that an integrated look at the process within the framework of the 
hydrokinetic model iHKM can lead to important insights.

In this note we, therefore, study the effects of hadronic rescatterings on $K^*$ resonance
observability within the integrated hydrokinetic model (iHKM) for the case of Pb+Pb
collisions at the Large Hadron Collider (LHC) energy
$\sqrt{s_{NN}}=2.76$~TeV and compare the results to the detailed data
recently reported by the ALICE collaboration~\cite{alice-k}. Our results 
are also compared with the STAR data from RHIC~\cite{kstar}.
 
\section{Integrated hydrokinetic model}

The 'Integrated Hydrokinetic Model' (iHKM)~\cite{ihkm} of relativistic
nuclear collisions is an extended version of the well-known HKM
model~\cite{HKM}, which includes, in addition to the latter, an
energy-momentum transport model of the pre-thermal stage of the matter
evolution~\cite{ihkm1,ihkm2} and viscous (not ideal as in HKM)
hydrodynamics at the thermal stage. As well as its predecessor, HKM, the iHKM model simultaneously describes well 
a wide class of bulk observables in heavy ion collisions, such as multiplicities, hadron transverse momentum spectra and their  anisotropy --- flow harmonics, interferometry radii, source functions, etc. at different centralities~\cite{ihkm, Uniform,Sf}.
Now the iHKM includes the five stages of the matter evolution and  observable formation in $A+A$ collisions:
   
	1) Formation of the initial state. 
	
	The estimates of the Glasma formation time (see \cite{ihkm} and references therein) as well as comparison of the iHKM results with the experimental data, point out on very early formation time  for initial energy density profile at the LHC, namely,  $\tau_0 \approx 0.1$ fm/$c$. To describe the energy density distribution at this initial state, the combined method is used~\cite{ihkm}. According to the latter, the generally non-equilibrium  boost-invariant (in the central region of rapidity) parton/gluon distribution function on the {\it initial} hypersurface $\sigma_0$: $\tau=\tau_0$ is presented in the factorized form 
	\begin{equation}
f(t_{\sigma_0},\textbf{r}_{\sigma_0},\textbf{p}) = \epsilon(b;\tau_0, {\bf r}_T)f_0(p).
\label{f0}
\end{equation}
Here, the initial energy density profile $\epsilon(b;\tau_0, {\bf r}_T)$ is calculated in hybrid approach, amending the so-called wounded nucleon model with the approach of binary collisions. The proportion between the two types of the corresponding contributions to the initial energy density is described by means of the phenomenological parameter, $0 \leq \alpha \leq 1$. The distributions of numbers of wounded nucleons and binary collisions at $\tau_0$ are calculated in \textsc{GLISSANDO}
code~\cite{Gliss}. The initial energy density profile defined then by the weighed (with the coefficients $\alpha$ and $1-\alpha$) sum of these distributions and the initial energy density $\epsilon_0$ in the center of the system, ${\bf x}_T\approx 0$, at the central collisions. They are  the main free parameters of the model. A possible momentum anisotropy of partons/gluons in the initial state, that is typical for approaches based on the Color Glass Condensate effective field theory, is taken into account by the function $f_0(p)$ in (\ref{f0}) in the way described in detail in Refs. \cite{ihkm, ihkm2}:
\begin{eqnarray}
f_0(p)=g \exp\left(-\sqrt{\frac{(p\cdot U)^2-(p\cdot
V)^2}{\lambda_{\perp}^2}+\frac{(p\cdot
V)^2}{\lambda_{\parallel}^2}}\right),
\label{anis1}
\end{eqnarray}
where $U^{\mu}=( \cosh\eta, 0, 0, \sinh\eta)$, $V^{\mu}=(\sinh\eta, 0, 0,\cosh\eta)$. 
In the rest frame of the fluid element, $\eta=0$, $(p\cdot U)^2-(p\cdot
V)^2=p_{\perp}^2$ and $(p\cdot
V)^2=p_{\parallel}^2$, so the parameters $\lambda_{\parallel}^2$ and $\lambda_{\perp}^2$ can be associated with the two temperatures along the beam axis and that orthogonal to it correspondingly. The parameter $\Lambda=\lambda_{\perp}/\lambda_{\parallel}$ thus defines the momentum anisotropy of the initial state. 

The values of  the initial transverse energy density $\epsilon_0$  at $\tau_0$
and ${\bf x}_T= 0$, and the parameter $\alpha$ are fixed based on the experimental mean charged particle
multiplicity values. It turns out that these two parameters do not
depend on the collision centrality at a given collision energy.  The
variations of other parameters, namely, the viscosity, relaxation time,
and thermalization time, lead to the re-scaling of the $\epsilon_0$ value (only!), necessary in order to re-adjust the model output to the experimental multiplicity~\cite{ihkm} in central collisions (see details in Ref. \cite{ihkm}). A change of the thermodynamic equation of state is correlated with corresponding modifications of the initial time and energy density.


2) Pre-thermal stage.

The non-termal energy-momentum tensor obtained from the distribution (\ref{f0}) is the starting point
for the subsequent pre-thermal dynamics in the relaxation time
approximation~\cite{ihkm,ihkm1,ihkm2} that results in thermalization of the
matter. The pre-thermal stage in the iHKM lasts optimally from the time of initial stage formation, $\tau_0 \approx 0.1$~fm/$c$, 
 till the thermalization time $\tau_{th}$, when the initially non-equilibrated system reaches an approximate local thermal equilibrium. In previous and this iHKM analysis we use the value $\tau_{th} = 1$~fm/$c$. 
The value of $\tau_0$ in the model is fixed based on the experimental pion $p_T$ spectrum slope. 
The temperature in the very central part of the system at the thermalization time $\tau_{th}$ is about 400~MeV at the LHC energy $\sqrt{s_{NN}}=2.76$~TeV, 
while the value average over transverse plane at this proper time $\tau_{th}$ is, of course, less.

3) The hydrodynamic stage.

At the thermalization time $\tau_{th}$ the energy-momentum tensor of the system takes the Israel-Stewart form 
for relativistic viscous hydrodynamics. The evolution of the system after time $\tau_{th}= 1$ fm/$c$ 
is described by the hydrodynamics with the minimal possible ratio of the shear viscosity coefficient 
to the entropy density $\frac{\eta}{s}=\frac{1}{4\pi}$. In this paper we use the two equations of state (EoS). 
One is that was used in HKM model. It is the Laine-Schroeder equation of state~\cite{EoS} that guarantees 
that the subsequent transition from the continuous medium to the hadron gas (with 360 hadrons being 
taken into account) will pass without discontinuities in pressure and energy density. 
The other one is taken from the recent Lattice QCD data \cite{lattice-hotqcd, lattice-wuppertal, EoS2}. 
The hydrodynamic approach is utilized to describe the expansion of superdense quark-gluon and hadron matter,
close to local chemical and thermal equilibrium, until the temperature at which both these types of
equilibrium are violated, and therefore another approximation should be used to describe the further system evolution. 

4) The particlization stage.

We assume that the chemically and thermally near-local-equilibrium evolution takes place until  certain temperature $T_p$. Then one performs a switching from hydrodynamic representation to the system description in terms of particles. In the iHKM either gradual or sudden particlization can be realized. 
In the current analysis we compare the two variants of sharp particlization in iHKM with the two temperatures: $T_p=163$~MeV that corresponds to the Laine-Shroeder equation of state for quark-gluon matter~\cite{EoS} and $T_p=156$~MeV responding to
HotQCD Collaboration equation of state~\cite{EoS2} (for briefness we will refer to it in what follows as ``HotQCD EoS''). 
Both switching temperatures are still in agreement with the Lattice QCD data for pseudo-critical temperature in the cross-over scenario, $T_{ps}=154\pm 9$~MeV~\cite{lattice-hotqcd, lattice-wuppertal}. 
The higher temperature, $T_p=163$~MeV, does not practically change our previous results~\cite{ihkm, HKM, Uniform} describing well the multitude of bulk observables at RHIC and LHC energies 
with the particlization temperature $T_p=165$~MeV. The latter corresponds to the energy density $\epsilon=0.5$~GeV/fm$^3$ for the Laine-Schroeder EoS, and serve as the reference point for comparison. The lower temperature, 
$T_p=156$~MeV, is considered as corresponding to the most recent estimates of chemical freeze-out temperature
in thermal model, $T_{ch}=156 \pm 1.5$ MeV~\cite{stachel-sqm2013}.
The particlization hypersurface is built in iHKM 
with the help of the Cornelius routine~\cite{Cornelius}. 
For taking into account the viscous corrections to the hadron distribution
function Grad’s 14-moment ansatz is used.
 
5) Hadronic cascade stage.

After particlization the system finally proceeds to the hadronic cascade within UrQMD model~\cite{urqmd}. In such type of the evolutionary model, the chemical freeze-out as well as the thermal one, are not sudden in time (and not sharp in temperature), in opposite to the thermal models, where the recent analysis points out to the temperature of sharp chemical freeze-out at the LHC energies $T_{ch}=156 \pm 1.5$ MeV~\cite{stachel-sqm2013}. From the point of view based on evolutionary hadron cascade models, the latter temperature should be considered rather as the effective temperature for continuous chemical freeze-out. The situation might be similar to the thermal continuous freeze-out. Namely, with regard to the latter, it is demonstrated \cite{continuous} that in good approximation there is the duality between descriptions of the particle spectra within realistic continuous freeze-out and within the Cooper-Frye prescription for sharp freeze-out, if the latter one is attributed to the hypersurface of maximal continuous emission of the hadrons from expanding fireball. The detailed analysis will be done in separate work; in this article we are basing on such a hypothesis of the duality between sharp and continuous chemical freeze-out, so here we consider the temperatures $T_p= 163$~MeV and $T_p=156$~MeV just as the particlization temperatures. At the further matter evolution the non-elastic processes (including annihilation but not the resonance decays) die out gradually so that the rate of decreasing of the inelastic collision number is maximal at some  effective ``chemical freeze-out'' temperature $T_{ch} \leq T_p$.    

In current calculations for the case of particlization temperature $T_p=163$~MeV we use the set of the iHKM parameters, which optimize the description of the multiple bulk observables~\cite{ihkm, Shap, PBM-Sin, mt_scales}, including circumscribing and prediction of pion and kaon interferometry radii momentum behavior, at the LHC: 
$\tau_0 = 0.1$~fm/$c$, $\tau_{th} = 1$~fm/$c$, the relaxation time at the pre-thermal stage $\tau_{rel} = 0.25$~fm/$c$, $\epsilon_0 = 680$~GeV/fm$^3$, $\alpha=0.24$, the momentum anisotropy of the initial state at $\tau_0$ --- the ratio of transverse to longitudinal ``temperatures'' is $\Lambda=100$, dissipative parameter $\eta/s=0.08$. As for the case of particlization temperature $T_p=156$~MeV, it differs from the previous one by the values of $\epsilon_0 = 495$~GeV/fm$^3$ and $\tau_0 = 0.15$~fm/$c$. 
The latter parameters guarantee the correct dependence of the multiplicity on the centrality and correct pion transverse momentum spectra for the corresponding ``HotQCD EoS''.

\section{Results and discussion}

The initial conditions (IC) for iHKM calculations are chosen to correspond to the simulation 
of Pb+Pb collisions at the LHC energy $\sqrt{s_{NN}}=2.76$~TeV
for the eight centrality classes: $c=0-5\%$, $c=5-10\%$, $c=10-20\%$,
$c=20-30\%$, $c=30-40\%$, $c=40-50\%$, $c=50-60\%$, and $c=60-70\%$.

\begin{figure}[t]
\centering
\includegraphics[bb=0 0 567 410, width=0.9\textwidth]{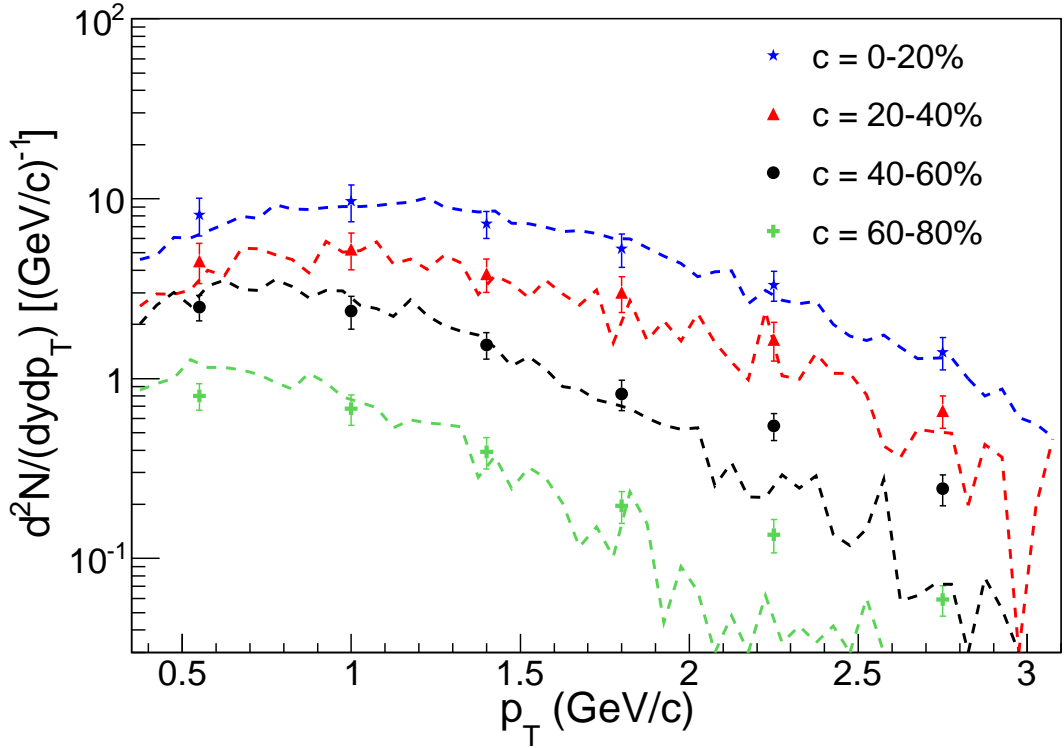}
\caption{The $K(892)^*$ resonance $p_T$ spectra for Pb+Pb collision 
events with different centralities at the LHC energy $\sqrt{s_{NN}}=2.76$~TeV
obtained in iHKM simulations (lines) in comparison with the experimental data~\cite{alice-k} (markers).
\label{kstars}}
\end{figure}

In Fig.~\ref{kstars} we present the comparison of $K(892)^*$ transverse
momentum spectra calculated in iHKM with the experimentally measured
ALICE points~\cite{alice-k}. The iHKM results are very close for both used EoS. One can see that iHKM reproduces well
the experimantal data for central collisions ($c=0-20\%$ and $c=20-40\%$ classes)
in full $p_T$ range and for the case of non-central collisions 
a good description is achieved for not very high $p_T<1.8$~GeV/$c$.

The $K(892)^{*0}$ resonances are identified in the experiment, as well as in iHKM, by means of the products
of their decays into $K^{+}\pi^{-}$ pairs. The result of such an identification is affected by  several factors. First, since strange $K(892)^*$ resonance has intermediate 
lifetime, about 4~fm/$c$, the intensive rescattering that takes place 
for particles born in course of hadronization, 
leads to the two opposed effects: a reduction of the number of $K\pi$ pairs (direct and coming from resonance decays)  identified as $K^*$
because of rescattering of mesons forming such a pair and,
on the other hand, an enhancement of identified $K^*$ because
of possible re-combination processes at the  ``afterburner'' stage. Apart from that, in experimental analysis,
different types of correlations, including event-by-event elliptic flow
and residual ones, which exist between kaons and pions, can be misinterpreted
as their bound resonance state. The common misidentification problem,
when particles of one species are identified as those of another one,
also contributes to this effect. And the last thing to note, the invariant mass criterion,
used to select $K\pi$ pairs corresponding to the $K^*$ decay, does not
work perfectly and can lead to rejecting some pairs that actually come from
the decay of interest or, on the contrary, to accepting some irrelevant pairs.

In order to investigate the influence of different above mentioned competing effects (except for experimental  
misidentification problem and event-by-event effects) on the $K^*$ observability, we use iHKM simulations.
Since the full iHKM calculation includes hadron cascade stage of system's evolution
modeled within UrQMD, one can compare the number of $K^{+}\pi^{-}$ pairs,
which can be identified as coming from the $K^{*0}$ decays after the UrQMD stage~\footnote{
Usually in the experiment one also analyzes $\overline{K^{*0}}$ and $K^{*\pm}$
yields, however in our simulations we found that all the results on efficiency of identification
for these resonances are very close to the results for $K^{*0}$, so in what follows
we will talk only about $K^{*0}$,  having in mind, that the same concerns also $\overline{K^{*0}}$ and $K^{*\pm}$.
} and the number of actual primary $K^{*0}$'s, produced in the course of medium particlization, 
plus those coming from subsequent resonance decays.
For our analysis we select the $K^{+}\pi^{-}$ pairs with rapidity $|y|<0.5$ and $0.3<k_T<5$~GeV/$c$ in accordance with \cite{alice-k}.
The criterion we use to tell that a pair comes from the desired decay
is the following: we require all spatial coordinates of the particle
last collision points to differ by less than 0.01~fm, $|x^{K}_i - x^{\pi}_i|<0.01$~fm,
and the pair invariant mass should fall in the range $0.77<M_{\pi K}<1.02$~GeV/$c^2$
around the $K(892)^{*0}$ invariant mass value, $M_{K^*}=895.94$~MeV, that corresponds
to the range utilized in the experimental analysis~\cite{alice-k}. 

\begin{figure}[t]
\centering
\includegraphics[bb=0 0 567 199, width=0.9\textwidth]{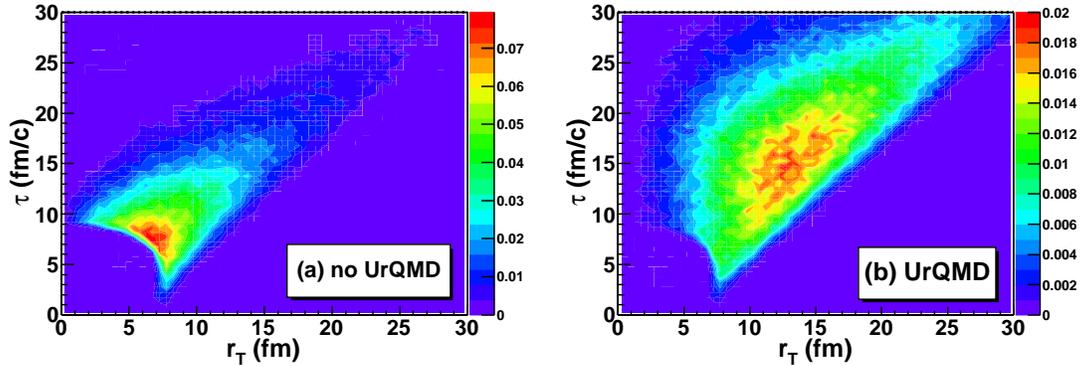} 
\caption{The comparison of the emission functions $ g(\tau,r_T)$, averaged over complementary 
space and momentum components, of $K^{+}\pi^{-}$ pairs, associated with $K(892)^{*0}$ decay products,
for two cases: (a) free-streaming of the particles and resonances, and (b) UrQMD hadron cascade.
The plots are obtained using iHKM simulations of Pb+Pb collisions at the LHC $\sqrt{s_{NN}}=2.76$~GeV, $0.3<k_T<5$~GeV/$c$, $|y|<0.5$, $c=5-10$\%.
\label{emiss}}
\end{figure}

Now, let us present the detailed analysis of the post-hydrodynamic stage of a central collision ($c=5-10$\%), 
based on the comparison of two scenarios after particlization: (1) free-streaming of the particles and resonances, 
and (2) UrQMD cascade. 
The emission pictures  of not re-scattered decay products of $K(892)^{*0}$ in both scenarios are presented in 
Fig.~\ref{emiss} for L.-S. EoS. We see that in the UrQMD scenario the emission picture is blurred out and strongly 
suppressed at small times after the particlization (see color scales). The numerical comparison of $K^{*}$ 
emission intensity at different proper times $\tau$ in two scenarios indicates that for the interval 
$\tau<15$~fm/$c$ in the UrQMD scenario the number of observed $K(892)^{*0}$'s is only 30\% of this number
in free-streaming case. Thus, at least 70\% of direct $K^{*}$'s cannot be detected due to rescattering of
their decay products in the dense hadronic medium, formed in the central collision. In fact, the actual loss
of direct $K^{*}$'s can be even greater because of $K^{+}\pi^{-}$ recombination taking place and producing
additional $K(892)^{*0}$ resonances -- in UrQMD it is approximated by the coalescence mechanism. 
This effect also could explain the picture we see at the large times, $15<\tau<30$~fm/$c$, 
where the number of observed $K^{*}$'s in the UrQMD case becomes about 1.5 times larger than that in the 
free-streaming case. 
If then one compares the total numbers of identified $K(892)^{*0}$'s in the two cases, it appears that the 
$K^{*0}$ reduction in UrQMD case is only about 20\% as compared to the calculation without rescatterings. 
This indicates that recombination plays important role in $K^{*}$ production and generates a great number of new 
$K^{*}$ particles, at least 50\% of direct ones, that partially compensate the substantial loss of the direct resonances taking place at small 
times after hadronization. Also this type of production can be the reason for comparably large time of maximal
emission, obtained for kaons in HKM study~\cite{mt_scales} and ALICE experimental analysis~\cite{Malinina}.

\begin{figure}[t]
\centering
\includegraphics[bb=0 0 567 409, width=0.9\textwidth]{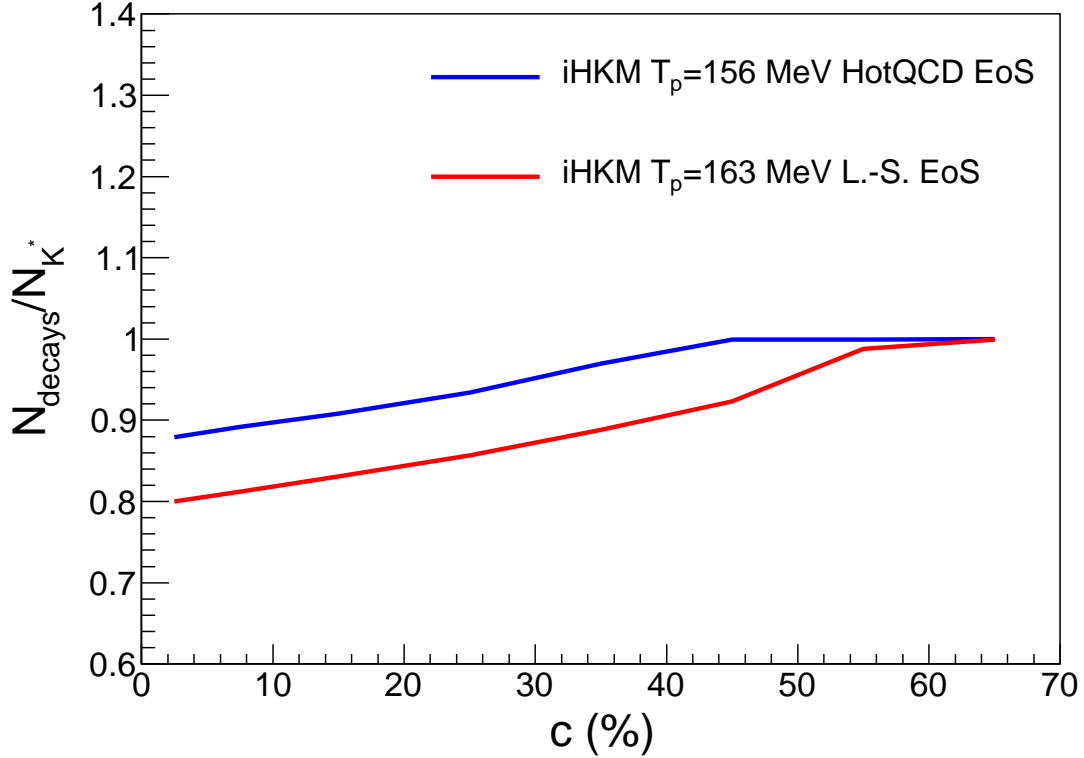} 
\caption{The fraction of $K^{+}\pi^{-}$ pairs coming from $K(892)^*$ decay,
which can be identified as daughters of $K^*$ in iHKM simulations after
the particle rescattering stage modeled within UrQMD hadron cascade.
The simulations correspond to LHC Pb+Pb collisions at $\sqrt{s_{NN}}=2.76$~TeV
with different centralities. The iHKM results are presented for two cases: the Laine-Shroeder equation of state
with particlization temperature $T_{p}=163$~MeV (red line) and the HotQCD equation of state with $T_{p}=156$~MeV (blue line).
\label{kstarf}}
\end{figure}

Now let us demonstrate the dependence of the fraction of identified $K^*$ resonances
on the Pb+Pb collision centrality at the LHC energy $\sqrt{s_{NN}}=2.76$~TeV, obtained in our analysis. 
Such dependencies are presented in Fig.~\ref{kstarf} for both EoS's and corresponding particlization temperature values. 
As one can see, at $T_{p}=163$~MeV this fraction increases from 0.8 for 5\% most
central collisions to unity for the near peripheral collisions with $c=60-70\%$.
In the case of $T_{p}=156$~MeV the fraction value also increases when going from central to peripheral
collisions, but for the central collisions it is greater, namely 0.88.
Such a behavior is expected, since at the periphery collisions most of colliding nuclei's
nucleons pass without intensive interaction and thus do not form large-volume
hot and dense long-living system, so $K^*$ decay products are not affected by re-scattering and can be easily identified.  
Also at higher particlization temperature one can expect longer duration of hadron re-scattering stage,
leading to more strong reduction of observable $K^{*}$ yield, as compared to that at the particlization stage,  due to its decay products' scattering. Note, as we will see further, that the final observed yields  of $K^{*0}$ are very similar at both EoS. 

As for the relatively longer living resonances, such as $\phi(1020)$ with lifetime about 50~fm/$c$
which decays into $K^{+}K^{-}$ and $K^0_L K^0_S$ pairs,
one cannot expect that the rescatterings of daughter particles will be noticeable to reduce observed resonance number. Since $KK$ interaction cross-section is not very large,
the recombination effect is also not big. The results of iHKM simulations are in
agreement with these considerations. 
The $\phi(1020)$ resonances were restored using $\phi \Rightarrow K^{+}K^{-}$ decay products 
selected based on the criterion $|x^{K^{+}}_i - x^{K^{-}}_i|<0.01$~fm
and $1.00<M_{K K}<1.07$~GeV/$c^2$, similar to the $K^{*}$ case (the invariant
mass range is taken from~\cite{alice-k}).
The fraction of observable $\phi(1020)$ in our simulations for all the centralities is about 20\% for non-central and 30\% for central collisions larger than unity. 
We connect such an enhancement with the manifestation of $KK$ correlations,
implemented in UrQMD through coalescence feature -- it transforms close $KK$ pairs into $\phi(1020)$ resonances.

\begin{figure}[t]
\centering
\includegraphics[bb=0 0 567 410, width=0.9\textwidth]{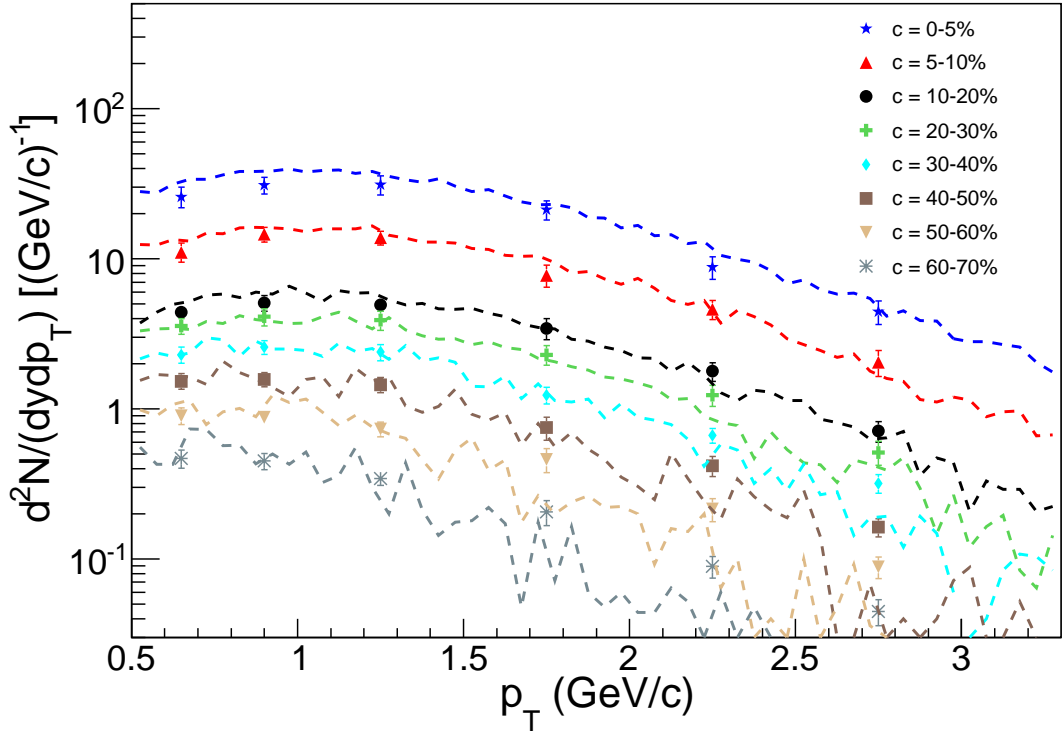}
\caption{The $\phi(1020)$ resonance $p_T$ spectra for Pb+Pb collision 
events with different centralities at the LHC energy $\sqrt{s_{NN}}=2.76$~TeV
obtained in iHKM simulations (lines) in comparison with the experimental data~\cite{alice-k} (markers).
\label{phis}}
\end{figure}

The $\phi(1020)$ $p_T$-spectra from iHKM simulations compared to
the ALICE data are presented in Fig.~\ref{phis}.
The $\phi$ mesons with rapidity $|y|<0.5$ and $0.5<p_T<5$~GeV/$c$ were
chosen for the analysis.
We observe the same situation as for $K^*$ spectra in Fig. \ref{kstars}-- the model
describes well the experimental spectra in the case of central events (up to $c=20-30\%$)
in full $p_T$ range, $0.5<p_T<5.0$~GeV/$c$, while for more peripheral
events model gives good description for the reduced range $0.5<p_T<1.8$~GeV/$c$.

In Tables~\ref{tab1}, \ref{tab2} one can  find the momentum integrated $K^{*0}$
and $\phi(1020)$ yields for different LHC collision centrality classes.
The results of iHKM simulations agree within the errors with the experiment.

\begin{figure}[t]
\centering
\includegraphics[bb=0 0 567 410, width=0.85\textwidth]{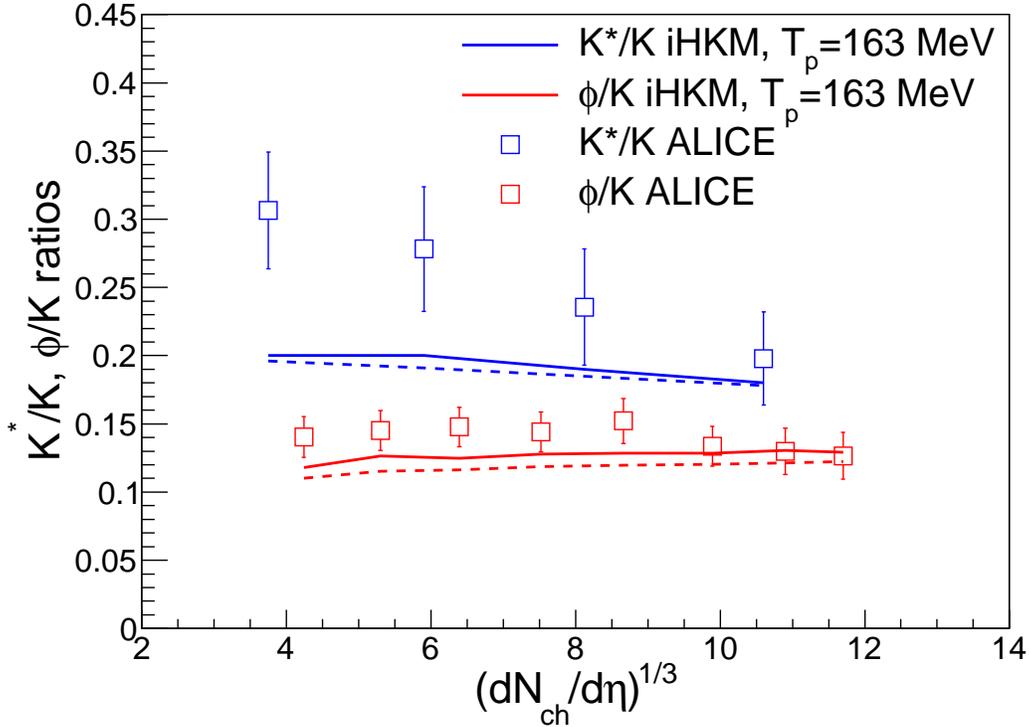} 
\caption{The comparison of $K^*/K^{+}$ and $\phi/K^{+}$ ratios dependency on
particle multiplicity $(dN_{ch}/d\eta)^{1/3}$ calculated in iHKM for the case of
LHC Pb+Pb collisions at $\sqrt{s_{NN}}=2.76$~TeV and the 
corresponding ALICE experimental data~\cite{alice-k}. The solid lines correspond to the iHKM calculations,
performed using hadronization temperature $T_{p}=163$~MeV and Laine-Shroeder equation of state~\cite{EoS}, while
the dashed lines are related to the HotQCD equation of state~\cite{EoS2} and $T_{p}=156$~MeV.}
\label{kstk-lhc}
\end{figure}

Additionally, we present the $K^*/K^{+}$ and $\phi/K^{+}$ particle number ratios,
calculated in the iHKM model for the LHC Pb+Pb collisions at the energy $\sqrt{s_{NN}}=2.76$~TeV
and for the top RHIC energy Au+Au collisions.
The results related to $T_{p}=163$~MeV are presented in Tables~\ref{tab1}, \ref{tab2}.
The corresponding values in the case of $T_{p}=156$~MeV for LHC are close to those in Table~\ref{tab1}
and are presented only graphically in Fig.~\ref{kstk-lhc}.

The centrality behavior of $K^*/K$ and $\phi/K$ ratios at the LHC, demonstrated in
Fig.~\ref{kstk-lhc} as the ratios' dependence on $(dN_{ch}/d\eta)^{1/3}$, is such that
the ``afterburner'' $K^*/K$ ratio slightly decreases at increasing collision centrality
(and multiplicity), while the $\phi/K$ ratio, on the contrary, slightly increases
with $(dN_{ch}/d\eta)^{1/3}$. 
The experimental $\phi/K$ points show not monotonic behavior --- at first the experimental $\phi/K$ ratio slightly increases, and then slightly decreases with the multiplicity. However, in general
one can say that the iHKM satisfactory describes the $\phi/K$ ratio and, for central events, also the $K^*/K$ ratio, but somehow underestimates the latter in case of the peripheral events. 

\begin{figure}[t]
\centering
\includegraphics[bb=0 0 567 410, width=0.85\textwidth]{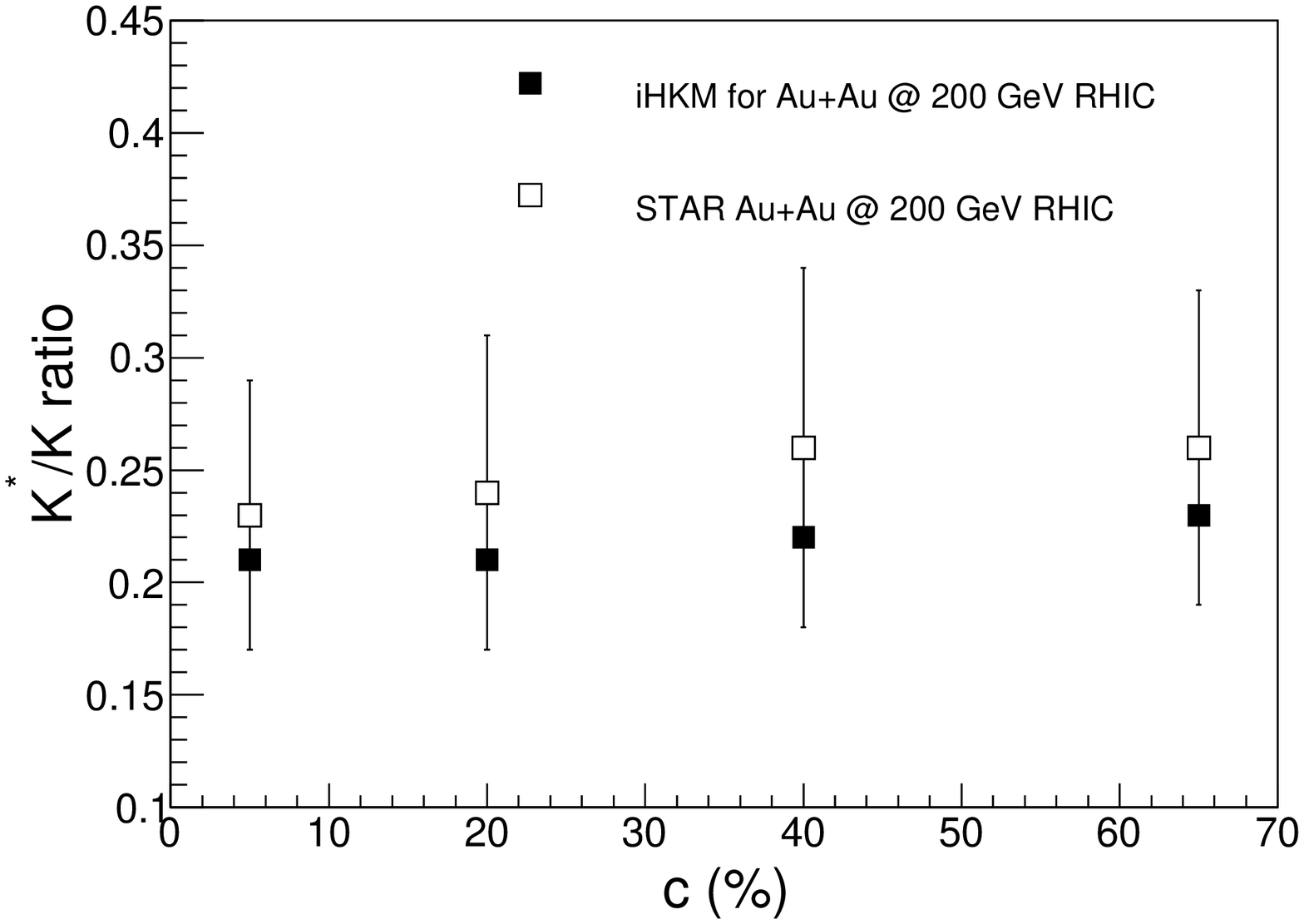}
\caption{The comparison of $K^*/K^{+}$ ratio calculated in iHKM for the case of
RHIC Au+Au collisions at $\sqrt{s_{NN}}=200$~GeV and the experimental data~\cite{kstar} for different centrality classes. }
\label{kstk}
\end{figure}

The results on $K^*/K$ ratio for RHIC are shown in Fig.~\ref{kstk} and are also compared with 
the corresponding experimental data~\cite{kstar}.
Here the ratio behavior is shown as dependence on collision centrality $c$.
Again, the $K^*/K$ ratio weakly decreases as the considered collisions become more central.
As one can see, our modeling results are in agreement with the RHIC experiment
within the errors. The presented experimental and model ratio values are related to the end 
of the collision's final stage. Within iHKM one can also compare them with the values on 
hadronization hypersurface, which for all the centralities appear to be approximately twice
larger than the ``afterburner'' ones.

The lower $K^*/K$ values for the central events and the absence of clear centrality
dependence for $\phi/K$ ratio are in agreement with above results concerning the $K^*$ and 
$\phi$ resonance observability --- in the $K^*$ case there is a reduction of its observed yield, 
which is the larger, the more central collision is considered, and in the $\phi$ case one has 
the enhancement of its observed number, which does not depend on centrality.
The thermal model results in the case of central LHC collisions~\cite{stachel-sqm2013} overestimate
the $K^{*}/K$ ratio by the factor of about 1.5. This fact supports the conception of continuous freeze-out,
realized inter alia through the particle interactions at the afterburner stage of collision.


\begin{table}
\begin{tabular}{ r c c c c }
\hline\hline
\multicolumn{5}{ c }{$K^{*}$}\\
Centrality & $dN/dy$ iHKM & $K^{*}/K$ iHKM & $dN/dy$ ALICE & $K^{*}/K$ ALICE \\\hline
0-20\% & 14.8 & 0.18 & $16.6 \pm 0.6 \pm 2.5 \pm 0.1$ & $0.20 \pm 0.01 \pm 0.03$ \\
20-40\% & 7.3 & 0.19 & $9.0 \pm 0.8 \pm 1.1 \pm 0.1$ & $0.24 \pm 0.02 \pm 0.03$ \\
40-60\% & 3.43 & 0.20 & $3.9 \pm 0.3 \pm 0.4 \pm 0.1$ & $0.28 \pm 0.02 \pm 0.03$  \\
60-80\% & 1.01 & 0.20 & $1.13 \pm 0.09 \pm 0.11 \pm0.07$ & $0.31 \pm 0.02 \pm 0.03$  \\\hline
\multicolumn{5}{ c }{}\\
\multicolumn{5}{ c }{ $\phi$ }\\
Centrality & $dN/dy$ iHKM & $\phi/K$ iHKM & $dN/dy$ ALICE & $\phi/K$ ALICE  \\\hline
0-5\% & 14.4 & 0.157 & $13.8 \pm 0.5 \pm 1.7 \pm 0.1$ & $0.127 \pm 0.004 \pm 0.014$  \\
5-10\% & 11.8 & 0.140 & $11.7 \pm 0.4 \pm 1.4 \pm 0.1$ & $0.130 \pm 0.004 \pm 0.014$  \\
10-20\%  & 8.6 & 0.135 & $9.0 \pm 0.2 \pm 1.0 \pm 0.1$ & $0.134 \pm 0.003 \pm 0.013$ \\
20-30\% & 5.6 & 0.131 & $7.0 \pm 0.1 \pm 0.8 \pm 0.1$ & $0.152 \pm 0.003 \pm 0.015$  \\
30-40\% & 3.56 & 0.128 & $4.28 \pm 0.09 \pm 0.48 \pm 0.09$ & $0.144 \pm 0.003 \pm 0.014$  \\
40-50\% & 2.06 & 0.125 & $2.67 \pm 0.05 \pm 0.30 \pm 0.06$ & $0.148 \pm 0.003 \pm 0.014$  \\
50-60\% & 1.11 & 0.126 & $1.49 \pm 0.03 \pm 0.16 \pm 0.05$ & $0.145 \pm 0.003 \pm 0.014$  \\
60-70\% & 0.52 & 0.118 & $0.72 \pm 0.02 \pm 0.08 \pm 0.04$ & $0.140 \pm 0.004 \pm 0.013$ \\\hline\hline
\end{tabular}
\caption{The $K^*$ and $\phi(1020)$ $p_T$-integrated yields, $K^*/K^{+}$ and 
$\phi/K^{+}$ ratios calculated in iHKM for the case of
LHC Pb+Pb collisions at $\sqrt{s_{NN}}=2.76$~TeV for the events from 
different centrality classes compared with the ALICE experimental data~\cite{alice-k}.}
\label{tab2}
\end{table}

\begin{table}
\begin{tabular}{|c|c|c|}
\hline
$c$ & $K^*/K$~STAR & $K^*/K$~iHKM \\ \hline
$0-10\%$   & $0.23 \pm 0.01 \pm 0.05$ & $0.21$\\ \hline
$10-30\%$  & $0.24 \pm 0.02 \pm 0.05$ & $0.21$\\ \hline
$30-50\%$  & $0.26 \pm 0.02 \pm 0.06$ & $0.22$ \\ \hline
$50-80\%$  & $0.26 \pm 0.02 \pm 0.05$ & $0.23$\\ \hline
\end{tabular}
\caption{The comparison of $K^*/K^{+}$ ratio calculated in iHKM for the case of
RHIC Au+Au collisions at $\sqrt{s_{NN}}=200$~GeV and the experimental data~\cite{kstar} for different 
centrality classes.} 
\label{tab1}
\end{table}

\section{Conclusions}

The production of strange $K^{*0}$ and long-lived $\phi(1020)$ 
(having hidden strange quark composition) resonances and the possibility for their reconstruction via the hadronic 
decay channels are considered within the integrated hydrokinetic model (iHKM) for the LHC and RHIC. It is found that the combined effect of the re-scattering and recombination of 
the daughter hadrons leads to the suppression of their registration up to 20\% in the most central collisions. 
The most intensive re-scattering takes place at times $\tau<15$~fm/$c$, where at least 70\% of direct
$K^{*}$ resonances become unobservable due to their decay products scattering. At larger times the recombination
effect becomes dominating leading to generation of a great amount of additional $K^{*}$'s, at least 50\% of direct ones.

The effectiveness of experimental $K(892)^*$ identification is significantly better for essentially non-central 
collisions, where the transverse size of the system is small and particles easily escape from the hadronic medium. 
As for $\phi(1020)$ resonance, it seems that $KK$ correlations at the afterburner stage lead to the excess of registered $\phi(1020)$. 
The calculations do not exclude that the full number of produced $\phi(1020)$ by up to 30\% 
exceeds their number on hadronization/particlization hypersurface. 
Such an effect, as well as additional $K(892)^*$ production at the final stage of the collision, 
can probably be explained by the regeneration of these resonances due to $K\overline{K}$ and $K^{+}\pi^{-}$ correlations at the afterburner stage through the UrQMD coalescence mechanism. 

The iHKM model with $T_{p}=163$~MeV reproduces the experimental data provided by the ALICE Collaboration
on $K^*$ and $\phi$ resonances transverse momentum spectra for $p_T<1.8$~GeV/$c$
in Pb+Pb collisions at the LHC energy $\sqrt{s_{NN}}=2.76$~TeV.
The results on $\phi/K$ and $K^{*}/K$ particle number ratio dependence
on particle multiplicity are described in iHKM sufficiently well, except for
$K^{*}/K$ ratio in the peripheral events. 
The corresponding $K^{*}/K$ values for RHIC numerically agree with the results of STAR 
Collaboration. 
The results on particle number ratios, obtained with $T_{p}=156$~MeV for the LHC are similar (only slightly worse),
to these at $T_{p}=163$~MeV. However, the comparison with the thermal model results points out
that the corrections connected with interactions at hadron afterburner stage of the matter evolution, are fairly noticeable. It supports the continuous freeze-out conception.

\begin{acknowledgments}
Yu.S. is grateful to ExtreMe Matter Institute EMMI/GSI for support and visiting professor position.  
The research was carried out within the scope of the EUREA:
European Ultra Relativistic Energies Agreement (European
Research Group: ``Heavy ions at ultrarelativistic energies''). 
The work is partially supported by Tomsk State University Competitiveness Improvement Program and the Physics and Astronomy Division of the NAS of Ukraine Program
``Structure and dynamics of statistical and quantum-field systems'' (PK № 0117U000240). 
\end{acknowledgments}

\end{document}